\documentclass[prd,a4paper,twocolumn]{revtex4}
\usepackage{graphicx}
 
\newcommand{\be}{\begin{equation}}
\newcommand{\ee}{\end{equation}}
\newcommand\lsim{\mathrel{\rlap{\lower4pt\hbox{\hskip1pt$\sim$}}
    \raise1pt\hbox{$<$}}}
\newcommand\gsim{\mathrel{\rlap{\lower4pt\hbox{\hskip1pt$\sim$}}
    \raise1pt\hbox{$>$}}}
\newcommand\esim{\mathrel{\rlap{\raise2pt\hbox{\hskip0pt$\sim$}}
    \lower1pt\hbox{$-$}}}

\begin{document}

\title{Scaling Laws for Non-Intercommuting Cosmic String Networks}

\author{C.J.A.P. Martins}
\email[Electronic address: ]{C.J.A.P.Martins@damtp.cam.ac.uk}
\affiliation{Centro de F\'{\i}sica do Porto, 
Rua do Campo Alegre 687, 4169-007, Porto, Portugal}
\affiliation{Department of Applied Mathematics and Theoretical
Physics, Centre for Mathematical Sciences,\\ University of
Cambridge, Wilberforce Road, Cambridge CB3 0WA, United Kingdom}
\date{20 September 2004} 
\begin{abstract} 
We study the evolution of non-interacting and entangled cosmic string networks in the context of the velocity-dependent one-scale model. Such networks may be formed in several contexts, including brane inflation. We show that the frozen network solution $L\propto a$, although generic, is only a transient one, and that the asymptotic solution is still $L\propto t$ as in the case of
ordinary (intercommuting) strings, although in the present context the universe
will usually be string-dominated. Thus the behaviour of two strings when they cross does not seem to affect their scaling laws, but only their densities relative to the background.
\end{abstract} 
\preprint{tba} 
\maketitle 
 
\section{Introduction} 
Cosmic strings \cite{VS} are topological defects that might have formed at phase transitions in the early universe. Interest in their evolution and cosmological consequences has recently grown, for both observational and theoretical reasons. On one hand, it is possible that a cosmic string has been identified as the source for an otherwise unexplained gravitational lens \cite{Sazhin1,Sazhin2}, or possibly even through a Kaiser-Stebbins effect in the cosmic microwave background. On the other hand, it has been recently realized that they can have a crucial role to play within superstring theory \cite{Copeland}, and in particular will be inevitably produced at the end of brane inflation \cite{Sarangi}.

Depending on the specific models, these cosmic strings may or may not have similar properties to the standard ones. An obvious difference is that they will appear in models with extra dimensions, though this alone does not guarantee that their evolution must be different. Another specific difference can potentially be more important, though. When two standard $U(1)$ strings interact, they always exchange partners (or intercommute), but in other specific models they may intercommute only with probability less than unity, pass through each other, or even entangle themselves via the formation of a bridge between the points of contact. It seems to be the case that these non-standard outcomes will be quite common in the higher-dimensional context, so it is interesting to ask what effect this will have, if any.

The evolution of standard cosmic strings has been extensively studied, both by
numerical \cite{bb,as,Moore,unified} and by analytic means \cite{Kibble,Kibble2,Bena1,Bena2,cka,ack,ms1a,ms1b,ms2b,extend}. Here one finds that after an initial transient period (whose duration depends on the string mass scale, being shorter for heavier strings) the network will evolve in a linear scaling regime, with a characteristic length (or correlation length) being a constant fraction of the horizon, $L\propto t$, and the RMS velocity being also constant.

Relatively little is known about the evolution of string with non-standard interactions. Relatively simple analytic arguments \cite{Dom2} suggest that they are frozen and conformally stretched ($L\propto a$) and end up dominating the energy density of the universe. The scenario of standard intercommuting strings dominating the universe was also discussed \cite{Dom1}, but it can never be realized for realistic networks. On the other hand, relatively small numerical simulations \cite{Vachaspati,Spergel,McGraw} report that linear scaling is reached except in special circumstances.

In this note we will use a suitably modified version of the velocity-dependent on-scale model of string evolution \cite{ms1a,extend} to clarify this issue. 
We will restrict ourselves to the standard case of three spatial dimensions, though we shall comment on the expected differences in the case of higher dimensions.
It will be shown that, just like in the standard case, an $L\propto a$ phase may exist but it will necessarily be a transient. Due to the significant string velocities, the network will tend to evolve towards a linear scaling regime, albeit one where the strings will usually be dominating the energy density of the universe.

\section{Analytic Modelling and Standard Linear Scaling}
The velocity-dependent one-scale model has been described in
detail elsewhere \cite{ms1b,extend},
so here we limit ourselves to a brief summary. Its underlying principle
is to replace a microscopic description of the string network
(provided say by the Goto-Nambu action) by a macroscopic one, based on a number of properties (such as a string correlation length $L$ and RMS velocity $v$)
whose evolution equations can be obtained by averaging the microscopic equations of motion. Any string
network divides fairly neatly into two distinct populations, long 
(or `infinite') strings and small closed loops.  
A phenomenological term must then be included to account for the loss of
energy from long strings by the production of loops, which are much smaller
than $L$---this is the `loop chopping efficiency' parameter $\tilde c$. One can 
obtain the following evolution equations
\begin{equation}
2\frac{dL}{dt}=2HL(1+v^2)+v^2\frac{L}{\ell_f}+{\tilde c}v\, , \label{evl}
\end{equation}
\begin{equation}
\frac{dv}{dt}=\left(1-{v^2}\right)
\left[\frac{k(v)}{L}-\left(2H+\frac{1}{\ell_f}\right)v\right]\, ;
\label{evv}
\end{equation}
where $H$ is the Hubble parameter and $k$ is a velocity-dependent parameter (called the `momentum parameter') which phenomenologically accounts for the presence of small-scale structures on the strings---see \cite{extend} for a thorough discussion. Both equations also contain friction term due to particle scattering, which is characterized by a friction length scale $\ell_f\sim\mu/T^3$.
This is only important in the earlier stages of the network evolution, and eventually becomes sub-dominant with respect to the damping from the expansion of the universe itself, so for simplicity we shall neglect it for the time being (though we will return to its effects later on in the paper).

Assuming that the scale factor behaves as $a\propto t^\alpha$ and defining $L=\gamma t$, the linear scaling solution is implicitly given by
\be\label{linscal}
\gamma^2=\frac{k(k+{\tilde c})}{4\alpha (1-\alpha)},\qquad v^2=\frac{1-\alpha}{\alpha}\frac{k}{(k+{\tilde c})}\,.
\ee
In the non-interacting limit ${\tilde c}\longrightarrow 0$, the requirement that the velocity is less than unity immediately implies that $\alpha>1/2$. Obviously this requirement applies for any value of ${\tilde c}$, which leads to the condition
\be\label{minc}
{\tilde c}>\frac{1-2\alpha}{\alpha}k\,.
\ee
So a linear scaling solution can exist in the matter era for an arbitrarily small loop chopping efficiency, but this is not the case in the radiation era (or any other epoch where the scale factor grows slower than that). This also shows that the role of loop production is quite different in the radiation and matter epochs, a point that has been noticed in numerical simulations \cite{bb,as}.

\section{Non-interacting Networks}
This case can be studied with our quantitative analytic model, simply taking ${\tilde c}=0$. Based on simple arguments \cite{Dom2} one expects the network to be frozen and conformally stretched, and indeed assuming a small velocity one trivially finds the $L\propto a$ solution. But a frozen network will eventually end up dominating the energy density of the universe, and when this happens, the scaling solution becomes
\be\label{noninov}
L\propto a \propto t,\qquad v=const.,
\ee
again assuming a small velocity, and with
\be\label{noninvgam}
\gamma^2=\frac{8\pi}{3}G\mu
\ee
So one could still say that the network is being conformally stretched, but this regime is physically very different from the previous one. The string domination has the effect of making the universe expand faster---in fact as fast as the strings themselves are allowed to by causality. As an aside, note that if standard intercommuting strings were to dominate the universe, $L\propto t$ and Eqn. (\ref{noninvgam}) would still hold, but the scale factor would evolve as $a\propto t^{2/3}$. Such a universe would therefore look like a matter-dominated one \cite{Dom1}, but as previously emphasized this is a scenario that can not be realized in practice.

There is, however, one factor that has been neglected in this analysis of non-interacting strings: the string velocities. Knowing how dynamically important the velocities are in the standard scenario \cite{ms1b,extend,unified} and even for non-relativistic strings in condensed matter \cite{extend,condmat} one could guess that they might have similar effects here, but these have not been studied so far.
Using the analytic model, it's easy to show that while the strings are still sub-dominant (and the scale factor is growing as $a\propto t^\alpha$) the scaling solution is
\be\label{vstretch}
L\propto a\,,\qquad v\propto t^{1-\alpha}\,.
\ee
So even though the correlation length is being conformally stretched and the velocities are small, they are in fact steadily increasing: the strings are being compressed and moving faster and faster. Even if string domination did not intervene to change the expansion rate, this scaling regime could only be transient. As an aside, we recall that a similar stretching regime exists for ordinary intercommuting strings, and in this case the scaling law for the string velocities is $v\propto t$. The effect of the intercommutings is to make the string velocities grow even faster (which is obvious, since it tends to introduce regions with higher curvature than average).

We can similarly generalize (\ref{noninov},\ref{noninvgam}) for the late-time, string-dominated epoch,
\be\label{noninovwithv}
L\propto a^{1+v^2}\,, \quad a\propto t^{1/1+v^2},\qquad v=const.,
\ee
with
\be\label{noninvgamwithv}
\gamma^2=\frac{8\pi}{3}G\mu\left(1+v^2\right)^2,
\ee
where the effects of the velocity correction are explicit. Again the strings are evolving as fast as allowed by causality ($L\propto t$). The non-negligible velocities make the evolution faster than conformal stretching, while making the universe expand more slowly because the string energy density is smaller (some of it being red-shifted away). Numerically  we find
\be
v_0^2\sim0.17\,,
\ee
so the scaling velocity, though smaller than the typical linear scaling velocities for intercommuting strings, is non-negligible. It is easy to show that this solution is an attractor for the analytic model.

\section{Entangled Networks}
Apart from intercommutation and non-interaction, there is a third possible outcome for the crossing of two cosmic strings: a bridge may form between them, at the point of the crossing. This leads to the so-called entangled networks
\cite{Vachaspati,Spergel,McGraw}. Again the naive expectation is that they might be conformally frozen, leading to the so-called frustrated networks.

In order to use our analytic model the evolution of these networks, we must extend it by including a term to account for the energy in the bridges. Since a segment of length $\ell$ moving with velocity $v$ has a probability $\ell v \delta t/L^2$ of crossing another string in the time $\delta t$, it is straightforward to calculate the total number of intercommuings in a given time and volume. Then, in the context of the approximations in the model, the energy density gained by the network as a result of the bridge formation is
\be\label{bridges}
\frac{d\rho}{dt}=\frac{v^2\rho t}{L^2}\,.
\ee
It is easy to see that at each moment most of the energy in this form has been produced in the previous Hubble time, so further effects are sub-dominant. One example is the annihilation of doubly-liked nodes, for which there is also some tentative numerical evidence \cite{Vachaspati,McGraw}. This term can therefore be included in the evolution equation for the correlation length (\ref{evl}). Within the context of a one-scale model, there is no further correction needed to the velocity equation (\ref{evv}). That this would not be the case in the context of more elaborate models where the string correlation length and curvature radius can be different.

We can now discuss the evolution of these networks. A noticeable point is that there is no conformal stretching solution: simple substitution in the evolution equations shows that if $a\propto t^\alpha$ and $L\propto a$, the only possible solution is $\alpha=1$ (in which case $v=const.$). Hence this solution is only possible if and when the strings dominate the universe.

In fact, early in the radiation era we find a much more dramatic transient solution
\be\label{tangrad1}
L=const.\,,\qquad v=\frac{L}{t}\,.
\ee
Such a solution exits only for a radiation-dominated epoch ($a\propto t^{/2}$), but not for any other behaviour of the scale factor. In the matter era, or indeed for any $a\propto t^\alpha$ with $\alpha > 1/2$, the solution is
\be\label{tangrad2}
L\propto t\,,\qquad v=const.\,,
\ee
but indeed this solution is reached much earlier than the normal epoch of radiation-matter equality, since with the transient solution (\ref{tangrad1})
the string network will very quickly dominate the universe, and the scale factor will then grow as $a\propto t$. Here the effect of the string velocities is negligible, unlike in the non-interacting case.
Linear scaling solutions seem to emerge from the simulations \cite{Vachaspati,Spergel,McGraw}. Unfortunately, none of these provides enough details of their results to enable a quantitative comparison (for example, none of them states whether their simulations are done in the radiation or matter epochs). It is hoped that future simulations can clarify this issue.

\section{Friction-dominated Networks}
So far we have neglected the effect of the frictional force due to particle scattering on the evolution of the networks. Since this will eventually become sub-dominant with respect to the damping from the expansion of the universe, with will not affect the asymptotic scaling laws. It could, however, affect some of the transients we have discussed.

Intercommuting strings in the expanding universe have two possible transient,
friction-dominated scaling solutions---see \cite{ms1b,extend} for derivations and discussion. If the network is formed with low initial density and velocity, it will initially evolve in brief stretching regime
\be\label{fricstr}
L\propto a\,,\qquad v\propto t\,.
\ee
As velocities increase will switch to the Kibble regime
\be\label{frickib}
L\propto t^{5/4}\,,\qquad v\propto t^{1/4}\,.
\ee
If the network is formed with high density, this solution will be immediately applicable. Finally when friction becomes sub-dominant and the network becomes relativistic the linear scaling regime is reached.

By analysing (\ref{evl}-\ref{evv}) it is easy to show that these transient solutions are still valid both for non-interacting and for entangled networks. This is obvious for the stretching regime if one 
remembers \cite{ms1b} that in this regime there is typically less than one string intersection per Hubble volume per Hubble time, so the outcome of the intersection should be immaterial. As a side remark, note that this is also true for the evolution of string networks during inflation, where
\be\label{infl}
L\propto a\,,\qquad v\propto a^{-1}\,.
\ee

Things are not that simple for the Kibble regime. However if one recalls (again see \cite{ms1b}) that the proportionality factors in (\ref{frickib}) involve the factor $(1+{\tilde c})$, then one sees that this will also be applicable for the case of non-interacting networks, the only difference being that the long string density and velocity at each epoch will be higher than in the standard case. This also turns out to be the case for the entangled networks, with an even higher string density.

Let us also comment on the evolution of these networks in the condensed matter context \cite{unified,condmat}. In this case there is obviously no expansion ($H=0$) but there is a constant friction force (whereas in the early universe this decays as $T^{-3}$). For standard intercommuting strings the asymptotic scaling laws are the well known
\be\label{condmatold}
L=\sqrt{(1+{\tilde c})}\left(\ell_f t\right)^{1/2}\,,\qquad v=\frac{k}{\sqrt{1+{\tilde c}}}\left(\frac{\ell_f}{t}\right)^{/2}\,;
\ee
and we can immediately see that they will still hold for non-interacting networks (${\tilde c}=0$), again with a higher string density and velocity. Just as in the case of the Kibble regime, this solution will also hold for entangled networks.

\section{Conclusions}
We have used the velocity-dependent one-scale model to quantitativel study the scaling behaviour of non-interacting and entangled string networks, thus clarifying some previously existing claims. 
We have shown that any $L\propto a$ scaling must be a transient, simply because of the effect of the increasing velocities, though usually this regime is ended when the strings dominate the energy density of the universe. When strings dominate, linear scaling ($L\propto t$, $v=const.$) is the attractor solution,
and the evolution of the scale factor of the universe depends on the string velocity. Usually this is negligible and it will grow as $a\propto t$, but for non-interacting strings the corrections are important---see (\ref{noninovwithv}). We have also shown that the transient scaling laws for friction-dominated evolution still hold for non-interacting and entangled networks, albeit with correspondingly higher string densities.

Although non-interacting and entangled cosmic string networks can be formed in a number of contexts in standard $(3+1)$-dimensional scenarios (see \cite{VS} for a review) the current wisdom seems to indicate that they should be more common  in higher dimensions. For example, two ordinarily intercommuting strings might miss each other when they cross if there are extra spatial dimensions. While it is beyond the scope of this note to analyse this case in any detail, we can nevertheless try to draw some lessons from the results above.

The main point is that $L\propto t$ scaling is a generic attractor: cosmic string networks will eventually be straightening out as fast as is allowed by causality. The background in which they find themselves and their own properties (what happens when two cross, their mass per unit length, and so on) can only influence the density and velocity at that linear scaling regime, and also what transient regimes might exist until linear scaling is reached. We expect that this will still be the case in higher dimensions. The main difference is likely to have to do with the fact that the sizes of any extra dimensions will be quite different from those of the three ordinary ones. One therefore expects that there will be anisotropies \cite{fossils} in the orientations and velocities of the strings, which could conceivably have observational consequences. Modelling such effects should require more than one correlation length and characteristic velocity. In that sense they would be somewhat analogous to models for wiggly \cite{wiggly} or superconducting \cite{supercond} strings.

\begin{acknowledgments}
I am grateful to Pedro Avelino, Anastasios Avgoustitdis and Paul Shellard for enlightening discussions on this topic, and to the hospitality of ICTP (Trieste) where this work was completed. 
\end{acknowledgments} 

\bibliography{strings} 
\end{document}